\documentclass[a4paper]{article}

\usepackage{icrc2013}
\usepackage[english]{babel} 

\title{Single-Mirror Small-Size Telescope Structure for the Cherenkov Telescope Array}

\shorttitle{1M-SST Structure for CTA}

\authors{
Jacek Niemiec,
Jerzy Micha\l{}owski,
Micha\l{} Dyrda,
Wojciech Kocha\'nski,
Jaromir Ludwin,
Marek Stodulski,
Pawe\l{} Zi\'o\l{}kowski,
Pawe\l{} \.Zychowski,
for the CTA Consortium.
}

\afiliations{
Institute of Nuclear Physics, Polish Academy of Sciences, Radzikowskiego 152, 31-342 Krak\'ow, Poland\\
}

\email{jacek.niemiec@ifj.edu.pl}

\abstract{A single-mirror small-size (1M-SST) Davies-Cotton telescope has been proposed for the southern observatory of the Cherenkov Telescope Array (CTA) by a consortium of scientific institutions from Poland, Switzerland, and Germany. The telescope has a 4 m diameter reflector and will be equipped with a fully digital camera based on Geiger avalanche photodiodes (APDs). Such a design is particularly interesting for CTA because it represents a very simple, reliable, and cheap solution for a SST. Here we present the design and the characteristics of the mechanical structure of the 1M-SST telescope and its drive system. We also discuss the results of a finite element method analysis in order to demonstrate the conformance of the design with the CTA specifications and scientific objectives. In addition, we report on the current status of the construction of a~aprototype telescope structure at the Institute of Nuclear Physics PAS in Krakow.}


\begin{document}
\maketitle

\section{Introduction}
A consortium of scientific institutions from Poland, Switzerland, and Germany has recently proposed a single-mirror small-size (1M-SST) telescope for the southern observatory of the Cherenkov Telescope Array (CTA) \cite{moderski}. The telescope utilizes a Davies-Cotton design -- a classic telescope type used in very high-energy ground-based astronomy. 1M-SST will be equipped with a fully digital and lightweight camera based on Geiger avalanche photodiodes (APDs) with the physical plate scale size of 24 mm (see \cite{boccone}). With the assumed angular pixel pitch 
$p = 0.25$ deg, the telescope focal length, i.e., the distance from the center of the dish to
the camera pixel plane, is thus $f = 5.6$ m. 
To assure that the optical point-spread function of the telescope is smaller than 0.25 deg at 4 deg off-axis, a 
reflective dish with diameter of 3.98 m is needed. This results in the camera field-of-view of 9 deg. The telescope's effective mirror area, corrected for the effect of shadowing by the camera and the mast, is 7.6 m$^2$. 

The 1M-SST telescope thus represents a compact structure. The use of silicon photomultipliers (PMTs) results in a low weight of the photodetector plane of 30 kg and the full camera weight of less than 300 kg. Hence, a number of engineering problems, challenged in earlier designs (e.g., H.E.S.S. or VERITAS telescopes \cite{hess,veritas}) are largely relaxed. The latter were caused by a necessity of supporting a heavy and large camera based on analog PMTs on the mast of a telescope with a large focal length ($f>10$ m). A~design of a low-cost solution for a 1M-SST that can be easily replicated to form a high-energy sub-array of the CTA is therefore feasible. 

The design of the 1M-SST structure is described in this work. In sections~\ref{telframe} and \ref{drive} the telescope's mechanical structure and drive system are presented, respectively. The results of the finite element method (FEM) analysis are shown in section~\ref{fem}.  
Finally, in section~\ref{summary} a current status of the construction of a prototype telescope structure at the Institute of Nuclear Physics PAS in Krakow is reported.

\section{Telescope Structure \label{structure}}
\subsection{Frame \label{telframe}}
\begin{figure*}[!t]
  \centering
  \includegraphics[trim=0.0cm 0cm 0cm 0.1cm, clip=true,width=0.59\textwidth]{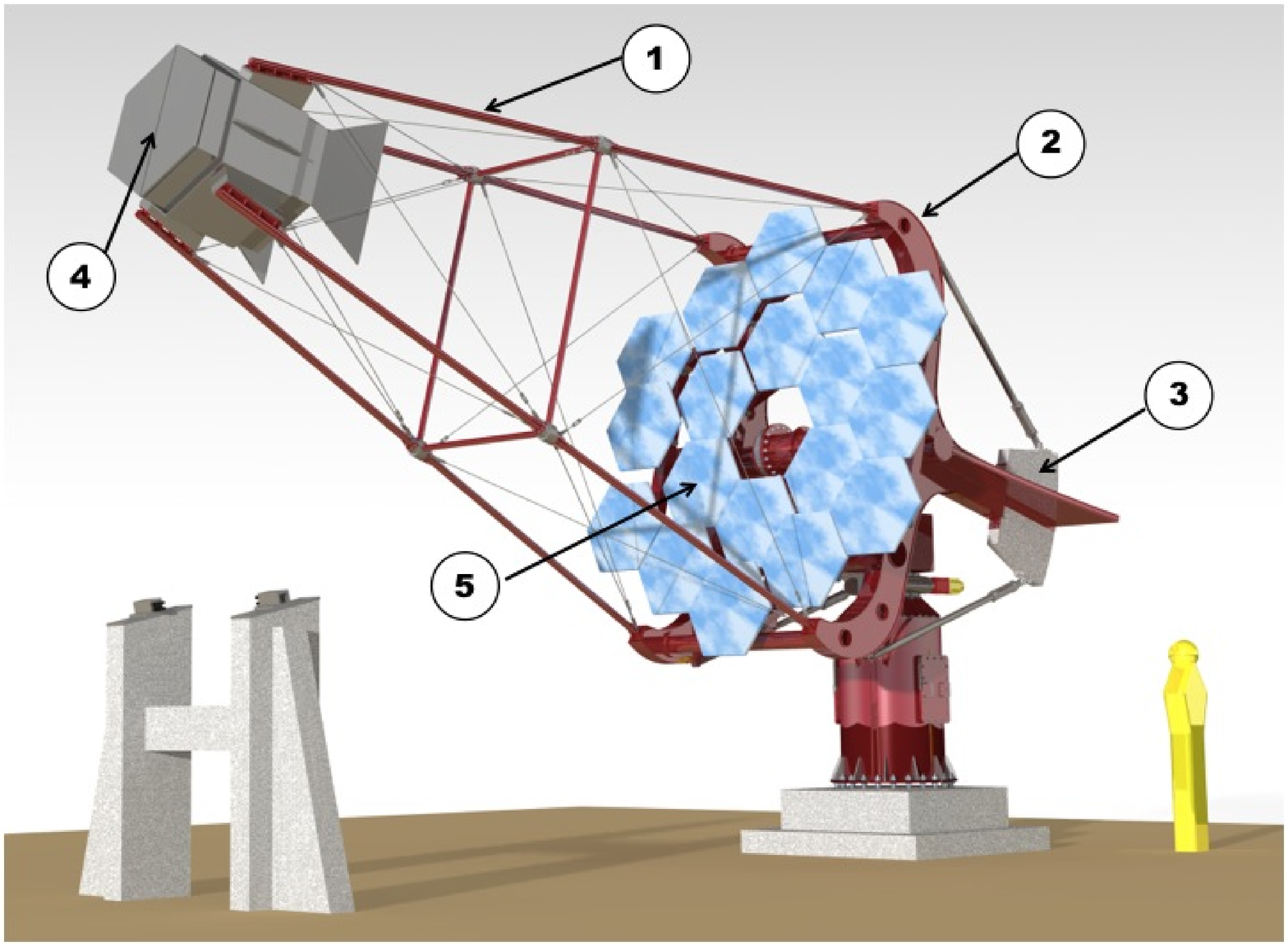}
  \includegraphics[trim=0.0cm 0cm 0cm 0.1cm, clip=true,width=0.394\textwidth]{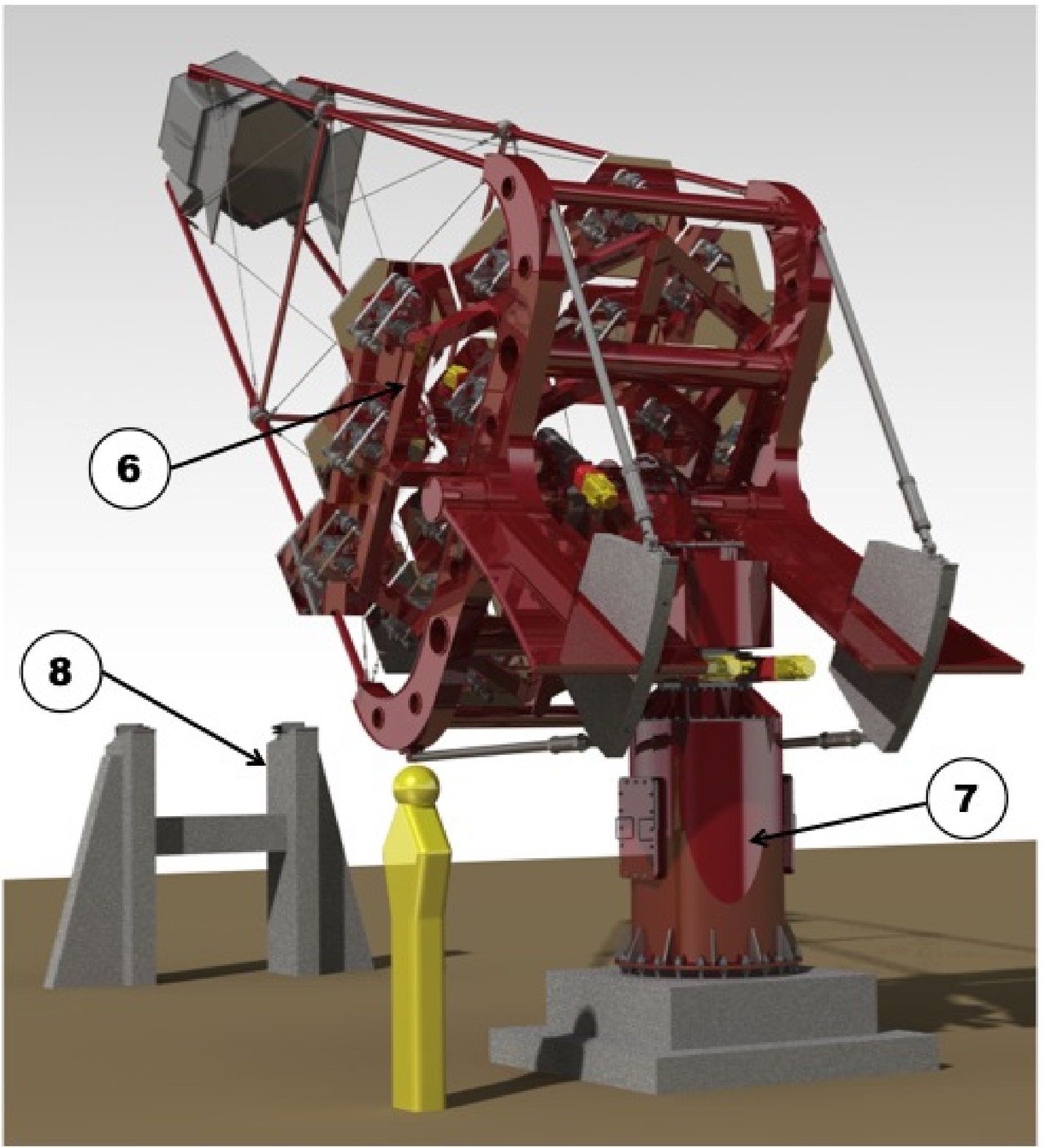}
  \caption{1M-SST sub-systems. Drawings in figures \ref{dish}, \ref{support}, and \ref{tower} are based on a technical documentation of the telescope structure. See text for the explanation of figure markings. }
  \label{telescope}
 \end{figure*}

The telescope frame consists of several sub-systems as shown in figure~\ref{telescope}. The mast (1) is directly connected to the dish support structure (2), to which the counterweight (3) is also attached. The rigidity of the mast is increased by the use of thin pre-tensioned steel rods. The mast positions the camera (4) with respect
to the mirrors (5) mounted on the dish (6). The latter is fixed to the dish support structure 
mounted on the telescope support (tower, 7).
Finally, the dock station (8) locks the telescope in a parking position.

The telescope is a compact and lightweight structure. Its hight is about 5 m, measured from the base of the tower, its width is 3.3 m, and the total length is 9.3 m. To cut the transportation costs a division of the telescope structure into smaller units that fit the standard size of a shipping container is applied. With steel used as a basic construction material, the telescope weighs less than 9 tons, including counterweights and the components of the drive system. The low weight, the fact that all steel profiles and tubes can be obtained as off-the-shelf products from the industry and that the design uses as many as possible identical
components lowers down the cost of manufacturing and guarantees a~low price of the telescope frame.

 \begin{figure}[!b]
  \centering
  \includegraphics[width=0.1755\textwidth]{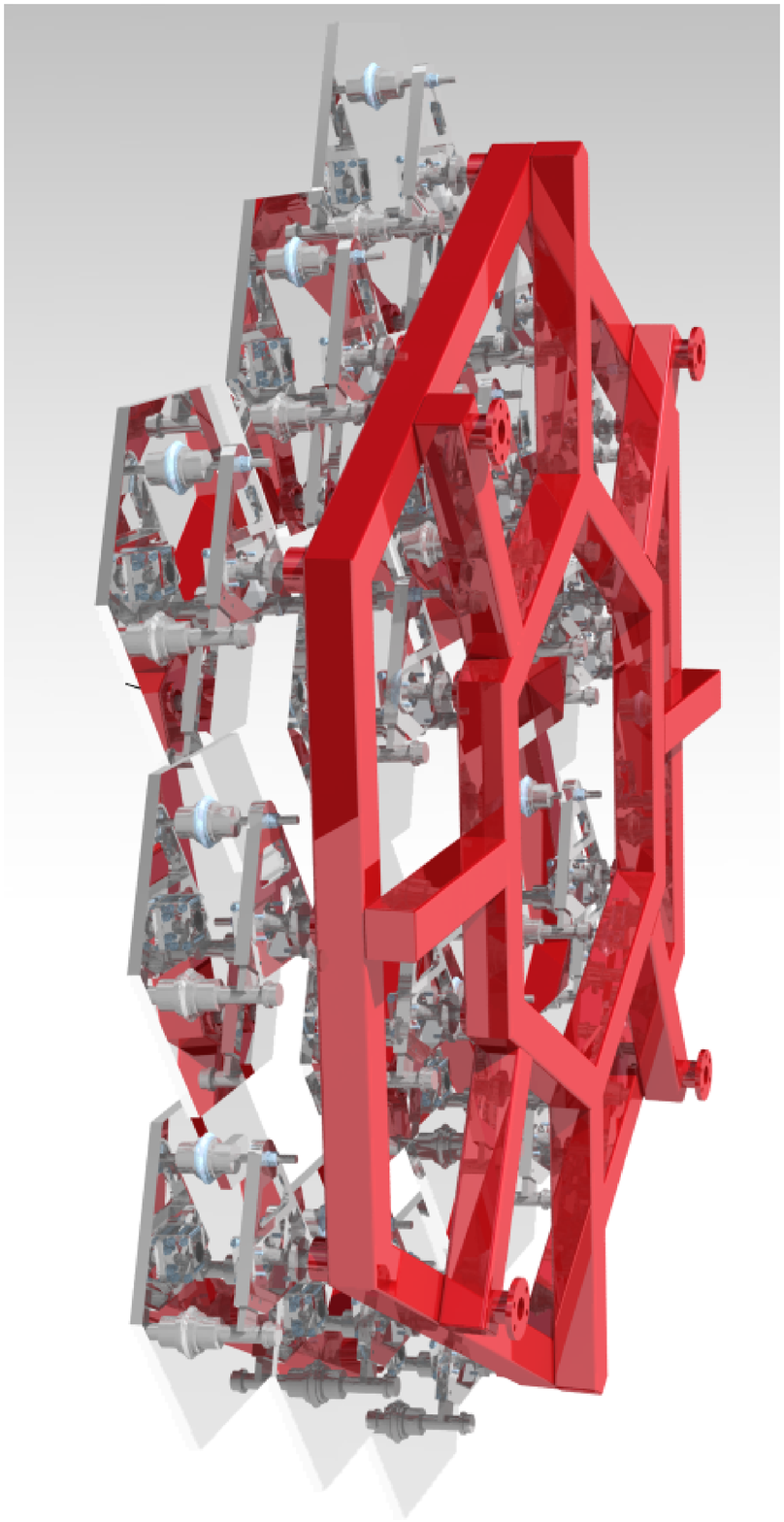}\hspace*{0.2cm}
  \includegraphics[width=0.28\textwidth]{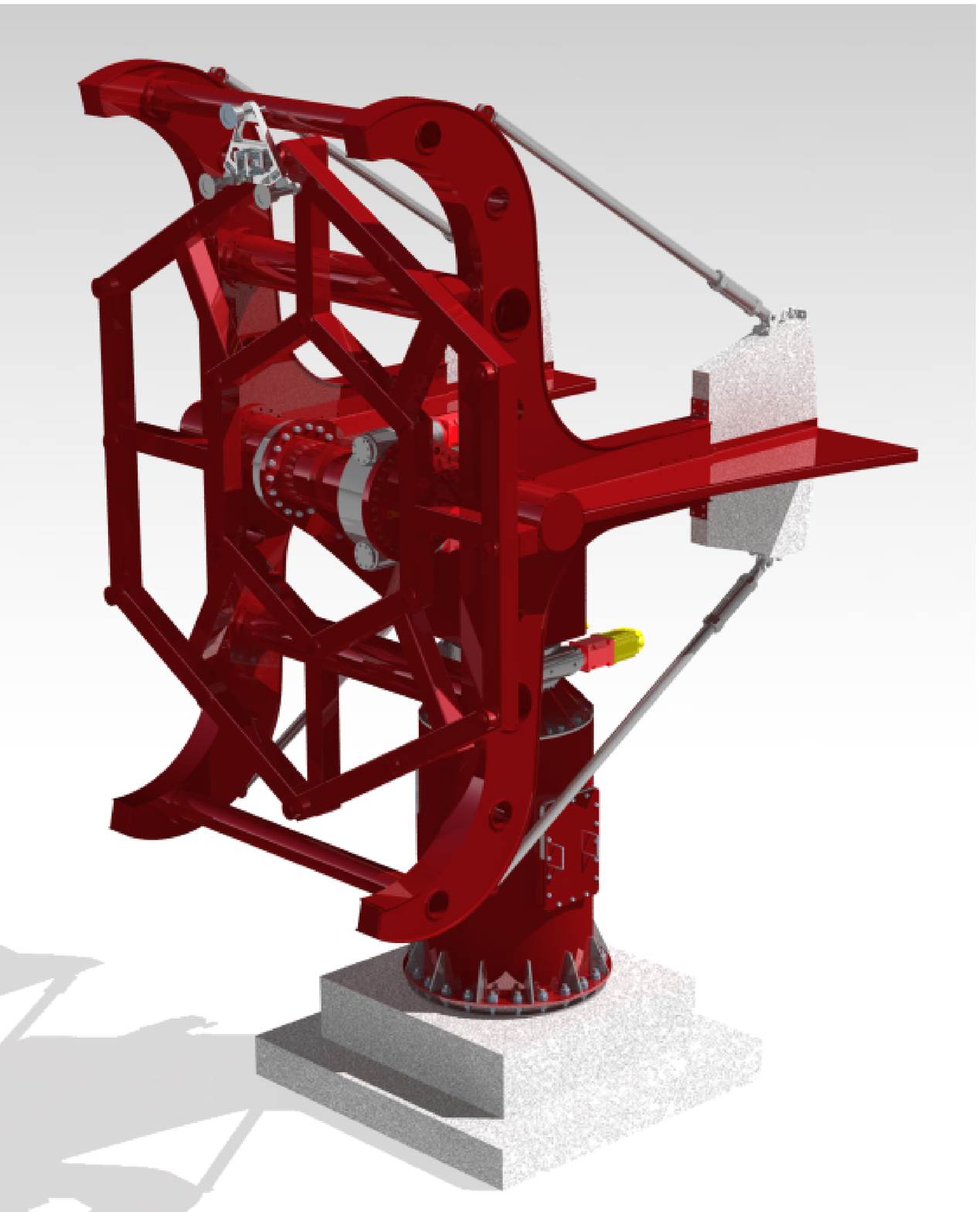}
  \caption{Layout of 18 mirror tiles on the dish in the rear/back view (left). The dish connected to the dish support structure (right).}
  \label{dish}
 \end{figure}

\begin{figure}[!b]
  \centering
  \includegraphics[width=0.27\textwidth,angle=-90]{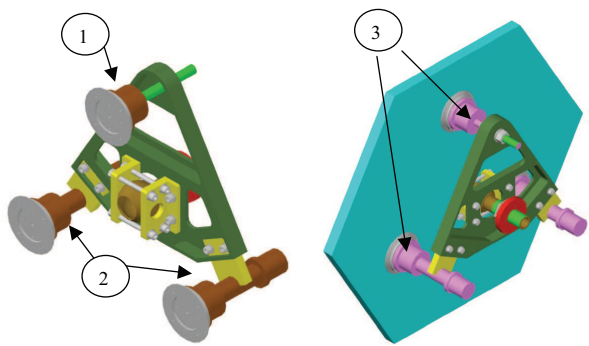}
  \caption{Mirror fixture with interfaces (left) and the mirror attached (right).}
  \label{mirror}
 \end{figure}

The layout of 18 mirror tiles that make up the reflecting dish
of 4 m diameter is shown in figure~\ref{dish} (left). Hexagonal mirrors of size 0.78 m flat-to-flat are used, with space between the tiles of 2 cm. The radius of curvature of a single mirror is 11.2 m. The dish structure is built of square steel profiles. It is to be connected to the dish support structure at four points, visible in figure~\ref{dish} as pads. The required spherical shape of the dish is obtained with two hexagonal sections welded to eight straight star-forming profiles. A precise positioning of the mirrors with respect to the camera is accomplished through a use of dedicated mirror fixtures (see figures~\ref{dish} and~\ref{mirror}). 

The mirror fixture design is presented in figure~\ref{mirror}. The fixture consists of an 
aluminium triangular frame with three fixing points and a steel threaded rod with a sphere at one end and a threaded flange at the other end. The flanges are bolted to the dish. The use of the 
threaded rod enables a precise positioning of a mirror at the focal length distance from the telescope's optical center. The sphere of the threaded rod is tightened in the central fixing
point of the triangular frame to create a hinge that allows for additional degrees of freedom during a~preadjustment of a~mirror orientation. 

The final precise alignment of the mirrors is to be performed by means 
of the automatic motor controllers (AMCs), called actuators. Each mirror fixture houses three interfaces that provide 3-point isostatic support to a mirror. One interface is a so-called 
fixed point support (1, figure~\ref{mirror}). 
Two other interfaces are actuators (2), which give a mirror facet two 
rotational degrees of freedom. 
The mirrors are fixed to these interfaces via steel pads (3) glued on the mirrors' rear surfaces.
Note, that the design of the mirror fixtures applied here does not only enable an arcminute-level stable alignment of the mirror facets relative to the camera, but it also allows for a simple and easy mounting/dismounting of the mirrors on/from the dish.

The design of the mast (see (1) in figure~\ref{telescope}) guarantees a proper setup of the camera with respect to the reflecting mirror surface at the telescope's focal distance of 5.6 m. The mast is built of eight circular tubes that are bolted together into four longer beams. The beams are connected with the camera housing at one end, and with the dish support structure at the other end. In this way the mast does not incur any mechanical stresses to the dish so that  
all mast deformations are practically disconnected from the deformations of the dish.
A rectangular frame roughly in the middle of the mast increases the stiffness of the mast structure and also facilitates its transport. The rigidity of the mast is additionally increased by the use of sixteen steel rods. The rods are pre-stressed with a force of 2000~N
and fixed to the long tubes by means of turnbuckles. 
To assure a proper geometry of the mast, eight special-design interfaces between the
rectangular frame and the beams must be used. Their shape is identical but requires computer numerical control (CNC) machining. Note, that such interfaces are not needed at points in which the mast is connected to the dish support structure.

 \begin{figure}[!t]
  \centering
  \includegraphics[width=0.46\textwidth]{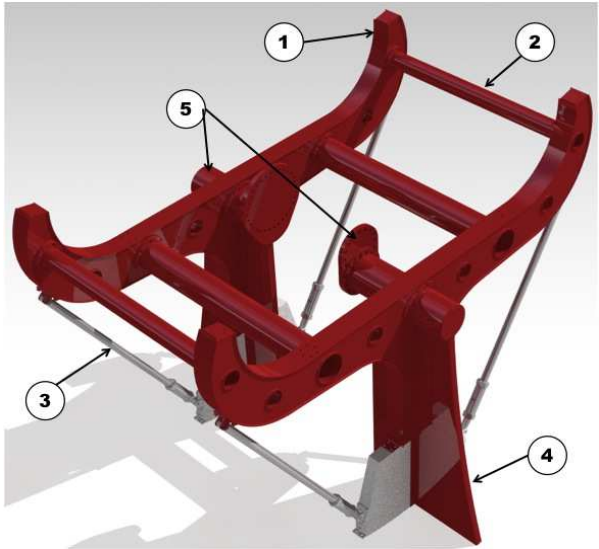}
  \caption{Design of the dish support structure.}
  \label{support}
 \end{figure}

 \begin{figure}[!b]
  \centering
  \includegraphics[width=0.47\textwidth]{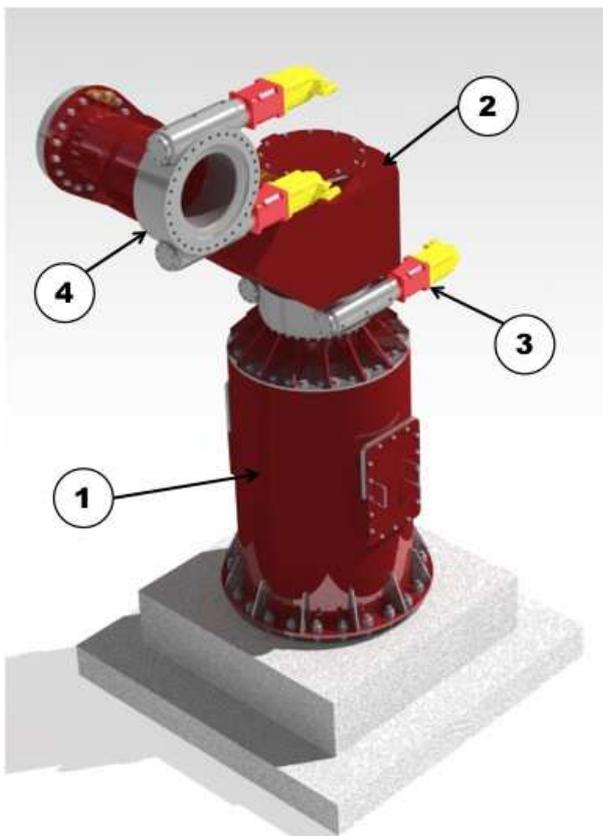}
  \caption{Design of the telescope support.}
  \label{tower}
 \end{figure}

The dish support structure is shown in figure~\ref{support}. It is composed of various
components. Two forks/Y-shaped frames (1) are box structures welded out of steel sheets and
tubes. The forks are spaced with four tubes (2). As mentioned above, the beams of
the mast are bolted to the four ends of the forks. Counterweights are fixed at their two
remaining ends. Four struts (3) and steel sheets (wings) (4) provide the necessary stiffness of the structure. The latter also improve the resistance to the wind blows. There are also two circular tubes with pads (5)
incorporated in the forks. The pads represent the interfaces to which the elevation drive system components (slew drive and the ball-bearing mounted on the sides of the head of the telescope support) are bolted.

The 1M-SST telescope support is built of two main components: the tower (1) and the head (2), figure~\ref{tower}.
The elements of the azimuth drive system (3) are incorporated into the tower, while these of the elevation drive system (4) into the head.
The tower must be bolted to a foundation in which 
a steel anchor grid structure with reinforcements is embedded. 
The slew drive (drawn in grey in figure~\ref{tower}) realizing the azimuth rotation of the telescope is fixed to the conical part of the tower (the tower cap). Inside the tower, there is a radial ball-bearing connected to the azimuth slew drive with the
tube, that stabilizes the whole structure. The housing of the azimuth ball bearing inside the tower can be accessed for installation and maintenance purposes via two openings.

The head is rotated by the azimuth slew drive. The elevation drive system components, including the slew drive and the ball-bearing, figure~\ref{tower}, are bolted to the pads of the dish support structure (see figure~\ref{support}). 
The slew drives and the bearings of both the azimuth and elevation drive systems are identical.

 \begin{figure}[!t]
  \centering
  \includegraphics[width=0.19\textwidth,angle=-90]{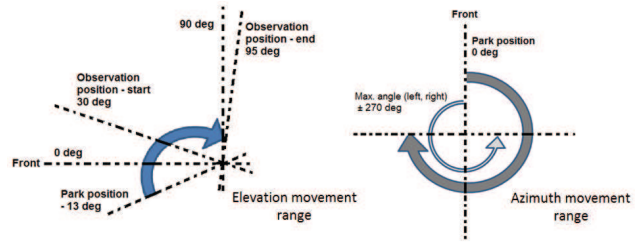}
  \caption{1M-SST drive system movement range in elevation (left) and azimuth (right).}
  \label{drive_range}
 \end{figure}

\subsection{Drive System \label{drive}}
As introduced in section~\ref{telframe}, a positioning and tracking system of the telescope is realized with two independent drive axes: the azimuth axis and the elevation axis.
The range for the telescope movement in elevation and azimuth is shown in 
figure~\ref{drive_range}.
The drive system of each axis is based on an
IMO \cite{imo} slew drive and a radial ball-bearing. An IMO slew drive is a compact system that combines a warm gear with a motor and also a roller bearing, thus enabling transmission of both radial and axial forces. It has a fully enclosed and self-supporting housing. 
Rotating components are fixed to the housing
with bolts. Each IMO slew drive selected for the 1M-SST contains a twin warm gear with two servo-motors. 
Such a solution helps to increase
the torque capacity and to eliminate backlash that is needed to achieve a required pointing accuracy of 7 arcsec and the tracking accuracy less than 5 arcmin. 
The drive system of the medium-size telescope (MST) for CTA is also based on the IMO slew drives and the choice of this type of drives for the 1M-SST has been partially done for the sake of the
uniformity among the various CTA telescope array sub-components. For this reason the concepts for control software and structure and safety will also be elaborated based on the MST solutions. 


\subsection{FEM Analysis \label{fem}}
\begin{figure}[ht]
  \centering
  \includegraphics[width=0.30\textwidth,angle=-90]{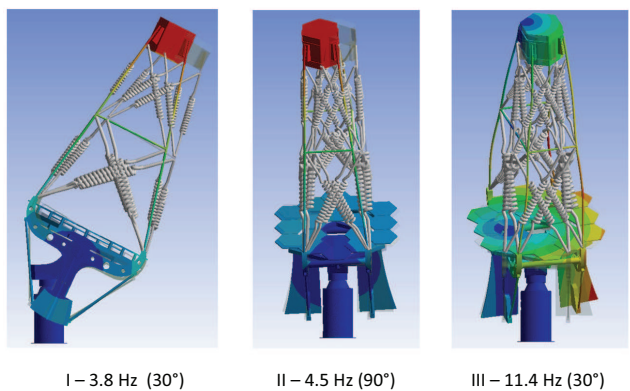}
  \caption{Results of the modal analysis showing the three lowest eigenfrequencies and their corresponding modes. The elevation angle is given in parantheses.}
  \label{modal}
 \end{figure}

The 1M-SST telescope structure design has been optimized based on the static FEM analysis. 
In order to perform such analysis a simplified computer-aided design (CAD) model of the telescope structure was built with ANSYS \cite{ansys}.
The model allows for an investigation of the structure deformations and mechanical stresses under gravity, wind, snow, and ice loads, and also for the earthquake conditions. 
The results show that the design of 1M-SST meets CTA optical requirement that the maximum
displacement of the camera with respect to the dish is smaller than 1/3 of the pixel size for the extreme observing mode conditions. The latter assume the maximum wind velocity with gusts of 50 km/h (the average velocity of 36 km/h) during which observation runs can still be taken. The resulting maximum displacement is 8.2 mm for the back-wind conditions and the telescope at 60 deg in the elevation angle. The maximum equivalent stresses in the structure are also well below the
elasticity limit for the steel. With $R_e=240$ MPa  the safety factor is about 3. Similar FEM analysis was conducted for the survival mode conditions. They allow for the wind velocity to reach 200 km/h while the telescope is locked in the parking
position. Also in this case all stresses are well below the plasticity of the materials used for the
mechanical construction. 
Preliminary FEM calculations show that the same is true if the earthquake loads are considered. 
Finally, the modal analysis, figure~\ref{modal}, demonstrates that the lowest eigenfrequency of the structure is 3.8 Hz. Thus it
is well above the minimum value of 2.5 Hz specified for CTA telescopes.

 Note, that the FEM calculations performed demonstrate a full conformance of the design with the CTA speciﬁcations and scientiﬁc 
objectives.

\section{Towards 1M-SST Prototype \label{summary}}
Based on the design presented in section~\ref{structure} a prototype telescope structure with drive system will be constructed at the Institute of Nuclear Physics PAS in Krak\'ow. The elements of the drive system have already been ordered. The technical design of the telescope frame has recently been prepared by an industrial partner. A call for tender for manufacturing of the telescope components and pre-installation at the INP PAS site will soon be placed. It is foreseen that the 1M-SST prototype will be ready for tests by the end of 2013.



\vspace*{0.5cm}
\footnotesize{{\bf Acknowledgment:}{This work was supported by The National Centre for Research and Development through project ERA-NET-ASPERA/01/10 and The National Science Centre through project DEC-2011/01/M/ST9/01891.
We also gratefully acknowledge support from the agencies and organizations
listed in this page: {\tt http://www.cta-observatory.org/?q=node/22}.}}

\end{document}